\documentclass[12pt]{article}
\usepackage{graphicx}
\usepackage{hyperref}
\pdfoutput=1
\usepackage{subfigure}
\usepackage{epsfig}
\usepackage{amsmath}
\usepackage{amsfonts}
\usepackage{amssymb}

\usepackage{amsmath,amssymb}
\usepackage{graphicx}
\usepackage{hyperref}
\usepackage{subfigure}
\usepackage{multirow}

\usepackage{color}
\newcommand{\dij}[1]{(\delta_{#1})_{ij}}
\newcommand{\dg}[2]{(\delta_{#1})_{#2}}

\newcommand{\EE}[1]{\times 10^{#1}}

\newcommand{\td}[1]{\tilde{#1}}

\newcommand{\eV}{{\rm eV}}

\newcommand{\GeV}{{\rm GeV}}
\newcommand{\TeV}{{\rm TeV}}
\newcommand{\ord}[1]{\mathcal{O}(#1)}

\newcommand{\sn}[1]{{\rm s}_{#1}}
\newcommand{\cn}[1]{{\rm c}_{#1}}

\def\[{\left [}
\def\]{\right ]}
\def\({\left (}
\def\){\right )}

\begin{document}

\begin{center}
{\bf \Large LFV in SUSY Breaking with Heavy Sleptons}
\end{center}

\vspace{0.02cm}
\begin{center}
{\sc Daniel Sword,\footnote{{\tt sword@physics.umn.edu}}$^a$} 
\end{center}

\begin{center}

$^a${\it\small School of Physics and Astronomy, 
University of Minnesota, Minneapolis, MN 55455, USA}

\end{center}

\vspace{0.2cm}

\begin{abstract}
We discuss lepton flavor violation (LFV) in the context of SUSY breaking when the first-two generation sleptons are heavy. When the first-two generation sleptons are $\mathcal{O}(5~{\rm TeV})$ and have small mass-splittings, we find that a light third generation slepton is allowed even with the large mixing implied by neutrino oscillation experiments. As an application, we consider a gravity dual to single-sector SUSY breaking and show that it is compatible with both LFV constraints and oscillation measurements.
\end{abstract}


\newpage

\section{Introduction}\label{intro}

A well-known consequence of many supersymmetric extensions of the standard model (SM)---notably those with non-universal soft supersymmetry (SUSY) breaking masses---is the appearance of flavor-violating processes at rates within reach of detection in present or near-future experiments. Much effort has been expended studying the various aspects of these phenomena in a wide variety of models. In particular, the study of lepton flavor violating (LFV) processes has been very active in the last several years, in part due to the enormous improvements in our understanding of neutrino physics. The compelling evidence in favor of neutrino oscillations (especially given the apparently large amount of mixing) has inspired a large number of papers discussing the implications for extensions of SUSY that incorporate neutrino masses.

The most common approach in previous studies has been in the context of seesaw-induced neutrino masses (and quite often some sort of GUT unification), as this sort of scenario offers an elegant mechanism for generating the tiny neutrino masses \cite{Altarelli:1999dg, Arganda:2007jw, Calibbi:2006nq, Masiero:2002jn, Casas:2001sr, Hisano:1998fj, Hisano:2442}. The specific results of any given analysis depend to some degree on the assumptions made regarding the texture of the neutrino Yukawa matrices. However, one important lesson one can take from these works (in particular, \cite{Casas:2001sr}) is that, in the absence of conspicuously convenient textures for the Yukawa matrices, the slepton masses must be very nearly universal.

One may avoid this sort of requirement if the superpartners happen to be very heavy, in which case the flavor changing processes are greatly suppressed \cite{Cohen:1996vb}. While the existence of heavy sparticles is seemingly at odds with one of the primary motivations of supersymmetry---namely, naturalness in the Higgs sector---non-universality actually allows for large first and second generation scalar masses without destabilizing the Higgs due to small Yukawa couplings \cite{Dimopoulos:1995mi}.

The purpose of this paper is to show how LFV constraints can be accomodated in this class of models when the first two generation sleptons are of $\mathcal{O}(5~{\rm TeV})$, even when one considers large mixing in the lepton sector as suggested by neutrino data. In fact, we find that when the first two-generation sleptons have a small splitting, the third-generation superpartner can in fact remain very light.

As an application, we consider a model proposed by Gabella, Gherghetta, and Giedt that features a similar spectrum \cite{Gabella:2007cp}. They describe a gravitational dual in five-dimensions to models of single-sector SUSY breaking. Generically, these models describe the first two generations of SM fermions as composite states of a strongly coupled gauge theory and the third generation as an elementary state. The superpartners of the composite states feel SUSY breaking directly and acquire large masses, whereas the elementary sfermions remain fairly light as they feel SUSY breaking only through gauge mediation \cite{Luty:1998vr}.

A major benefit of 5D dual models is that they allow for a quantitative calculation of the 4D mass spectrum. While the SM particle masses arise from wavefunction overlap with a brane-localized Higgs field after integrating over the extra dimension, a ``warp factor'' can be used to set the scale of dynamical SUSY breaking. Moreover, a very large range of Yukawa couplings---corresponding to the top quark mass all the way down to tiny Dirac neutrino masses---can be quite naturally explained in terms of order unity localization parameters.\footnote{This sort of mechanism for generating Dirac neutrino masses has been discussed, e.g., in \cite{Grossman:1999ra, Gherghetta:2003he}}

The layout of this paper is as follows: In section \ref{neutrinos}, we will briefly review the physics of neutrino oscillations. In section \ref{flavor}, we examine the existing flavor constraints on models with hierarchical soft terms. We find that LFV constraints can be accomodated if the mass splitting between the first two generations is not too large. In section \ref{dual}, we review the setup of the model described in Ref. \cite{Gabella:2007cp}. We then apply the results of our analysis to this model and adjust the model to respect bounds from flavor physics. Our full results are presented after two-loop numerical RGE from the messenger scale to the electroweak scale.


\section{Neutrino Oscillations}\label{neutrinos}


\begin{table}[ht]
\begin{center}
\begin{tabular}{|| c | c ||}
\hline\hline
Parameter & Best Fit\\
\hline
$\rm{sin}^{2}\theta_{12} $ & 0.32\\
$\rm{sin}^{2}\theta_{23} $ & 0.50\\
$\rm{sin}^{2}\theta_{13} $ & 0.007\\
$\Delta m_{12}^2$ ($10^{-5}~\eV^2$) & 7.6\\
$\Delta m_{23}^2$ ($10^{-3}~\eV^2$) & 2.4\\
\hline\hline
\end{tabular}
\caption{Best-fit values for the neutrino mixing angles as they appear in Ref. \cite{Maltoni:2004ei}.\label{mixingangles}}
\end{center}
\end{table}

The introduction of neutrino mass to the MSSM is necessary in order to account for neutrino oscillations. The most recent neutrino oscillation data has been analyzed and best-fit values for the mixing angles and mass splittings have been determined in Ref. \cite{Maltoni:2004ei}. These are reproduced in Table \ref{mixingangles}. In addition, there is a limit on the mass of the heaviest neutrino obtained from cosmological data, $m_{\nu}^{\rm max}<0.7~\eV$ \cite{Amsler:2008zzb}. There remains an ambiguity in the neutrino mass hierarchy \cite{DeGouvea:2005gd}, in that it is unknown whether the mass eigenstates align in a ``normal'' or ``inverted'' way. In the normal (inverted) hierarchy picture, the lightest neutrino mass eigenstate is most closely aligned with the electron (tau) neutrino flavor eigenstate.

The neutrino and charged lepton Yukawa matrices in the weak eigenbasis, $Y_{e/\nu}$, may be related to those in the mass eigenbasis, $m_{e/\nu}$, by a biunitary transformation as:
\begin{eqnarray}
\label{rotationmatrices}
v_u Y_{\nu}=U_{\nu_L}^{\dag}m_{\nu}^{(d)}U_{\nu_R}\nonumber\\
v_d Y_{e}=U_{e_L}^{\dag}m_{e}^{(d)}U_{e_R}.
\end{eqnarray}
The product of the two left-handed rotation matrices is constrained as:
\begin{equation}
\label{pmnsconstraint}
U_{e_L}^{\dag}U_{\nu_L}=U_{PMNS},
\end{equation}
where $U_{PMNS}$ is the Pontecorvo-Maki-Nakagawa-Sakata matrix. From the values of Table \ref{mixingangles}, one calculates the entries of the PMNS matrix to be:
\begin{equation}
\label{pmnsmatrix}
U_{PMNS}=\( \begin{array}{ccc}
\cn{12}\cn{13}&\sn{12}\cn{13}&\sn{13}\\
-\sn{12}\cn{23}-\cn{12}\sn{23}\sn{13}&\cn{12}\cn{23}-\sn{12}\sn{23}\sn{13}&\sn{23}\cn{13}\\
\sn{12}\sn{23}-\cn{12}\cn{23}\sn{13}&-\cn{12}\sn{23}-\sn{12}\cn{23}\sn{13}&\cn{23}\cn{13}
\end{array} \)=\( \begin{array}{ccc}
0.822&0.564&0.084\\
-0.449&0.550&0.705\\
0.351&-0.617&0.705
\end{array} \).\nonumber\\
\end{equation}
We employ the notation $\cn{ab}=\cos \theta_{ab}$ ($\sn{ab}=\sin \theta_{ab}$). Furthermore, we have ignored the Majorana phases, $\alpha_1$ and $\alpha_2$, since we will be assuming Dirac-type neutrino masses only, as well as the CP-violating phase, since it is unconstrained.


There is still considerable freedom in choosing the four rotation matrices, as the measured mixing angles only constrain the combination of the two matrices. For example, one can consider the possibility that the matrix $U_{\nu_L}$ is the unit matrix and the mixing is among the charged leptons only. Such an approach generically leads to difficulty in explaining the smallness of $\theta_{13}$, however \cite{Altarelli:2004jb}. Moreover, we find sufficient LFV suppression under such an assumption generically requires a heavy third generation slepton in addition to the first two. Therefore, we will adopt the following assumption for simplicity:
\begin{eqnarray}
v_u Y_{\nu}=U^T m_{\nu}^{(d)} U\label{assumption},\\
\end{eqnarray}
with $U=U_{PMNS}$, and from here on we drop the ``PMNS'' label.

\begin{figure}
\centerline{\includegraphics{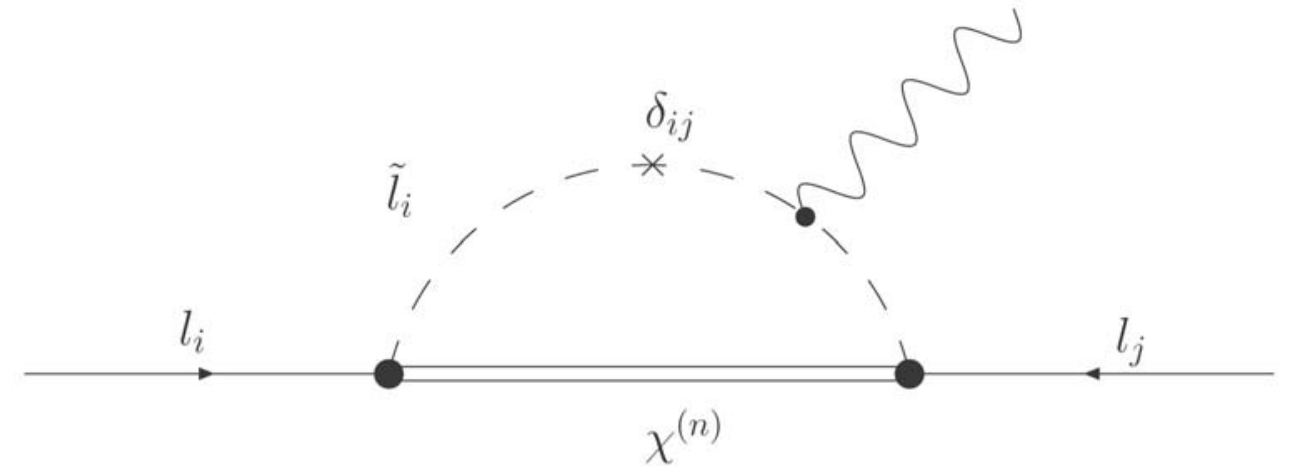}}
\caption{Generic Feynman graph leading to lepton decay. In addition to loops involving neutralinos, there are also processes involving charginos and sneutrinos, as well as those involving higgsinos. Each process involves a chirality flip (indicated by the change in arrow directions), which may take place on either the external or internal lines or, in higgsino loops, at a vertex (c.f. \cite{Hisano:1998fj}).\label{feynmangraph}}
\end{figure}



\section{LFV in the MSSM}\label{flavor}
\subsection{The SCKM Basis\label{sec:SCKM}}

The supersymmetric contributions to flavor changing amplitudes are typically calculated in the mass insertion approximation (MIA), in which the branching ratios are expanded in terms of small parameters, $\dij{AB}$. These parameters measure the size of off-diagonal entries in the slepton mass matrices when expressed in the lepton mass eigenbasis. In this basis, known as the Super-CKM (SCKM) basis, the slepton mass matrices take the form of a $6\times 6$ matrix \cite{Hisano:2442}: 
\begin{eqnarray}
\label{sleptonmassmatrix}
m_{\td{f},SCKM}^{2}&=&\( \begin{array}{cc}
(m_{\td{f}}^2)_{LL} + m_f^2 + M_Z^2 \cn{2\beta}(T_{f_L}^{(3)}-Q\sn{W}^2)& (m_{\td{f}}^2)_{LR} \\
(m_{\td{f}}^2)_{LR}^{\dag} & (m_{\td{f}}^2)_{RR} + m_f^2 - M_Z^2 \cn{2\beta}(T_{f_R}^{(3)}-Q\sn{W}^2)\end{array} \)\nonumber\\
&\equiv&\( \begin{array}{cc}
(M_{\tilde{f}}^{2})_{LL}& (M_{\tilde{f}}^{2})_{LR} \\
(M_{\tilde{f}}^{2})_{RL} & (M_{\tilde{f}}^{2})_{RR}\end{array} \).
\end{eqnarray}
Here $T_{f_{L,R}}^{(3)}$ is the third component of weak isospin, $f=\nu,l$ for sneutrinos or charged sleptons, respectively, and we employ the notation $\sn{W}\equiv\sin \theta_W$ ($\cn{W}\equiv\cos \theta_W$) where $\theta_W$ is the Weinberg angle. The $3\times 3$ matrices $(m_{\td{f}})^2_{AB}$ are given by
\begin{eqnarray}
\label{flavorchangingentries}
(m_{\td{f}}^2)_{LL/RR} &=& U_{f_{L/R}} m_{\td{f}_{L/R}}^2 U^{\dag}_{f_{L/R}}\nonumber\\
(m_{\td{f}}^2)_{LR} &=& \Bigg{\{}\begin{array}{l} v_u U_{f_L}A_{\td{f}} U_{f_R}^{\dag}-\mu m_f \cot \beta,\qquad f=\nu\\
v_d U_{f_L}A_{\td{f}} U_{f_R}^{\dag} -\mu m_f \tan\beta,\qquad f=l
\end{array}
\end{eqnarray}
with $\mu$ being the usual supersymmetric Higgs mass parameter and $A_{\td{f}}$ the matrix of soft trilinear scalar couplings. From \eqref{sleptonmassmatrix}, we calculate the $\dij{AB}$'s as (suppressing the slepton type label, $\td{f}$):
\begin{equation}
\dij{AB}\equiv\frac{(\Delta_{AB})_{ij}}{\sqrt{\(M^{2}_{AA}\)_{ii}\(M^{2}_{BB}\)_{jj}}}=\frac{\(M^{2}_{AB}\)_{ij}}{\sqrt{\(M^{2}_{AA}\)_{ii}\(M^{2}_{BB}\)_{jj}}}.
\label{dijforma}
\end{equation}
The notation $M$ here indicate the inclusion of the electroweak symmetry breaking (EWSB) contributions to the slepton mass matrices, as in \eqref{sleptonmassmatrix}. The EWSB contributions to the diagonal entries of the matrices $M^{2}_{AB}$ are small compared to the soft masses, as are the radiative corrections to the matrices $m_{{L/R}}^2$. To within a few percent, we can estimate the size of the $\dij{LL/RR}$'s as
\begin{equation}
\dij{LL/RR}\simeq\frac{\(U_{f_{L/R}} m_{f_{L/R}}^2 U^{\dag}_{f_{L/R}}\)_{ij}}{\sqrt{\(U_{f_{L/R}} m_{f_{L/R}}^2 U^{\dag}_{f_{L/R}}\)_{ii}\(U_{f_{L/R}} m_{f_{L/R}}^2 U^{\dag}_{f_{L/R}}\)_{jj}}},
\label{dijform}
\end{equation}
where the $m_{L/R}^2$ may be taken as the diagonal matrices of soft input masses. Note that we use these results simply to estimate the allowed parameter space, whereas a full numerical two-loop calculation will be used for application to the model in the following section.




\subsection{Constraints\label{fcncconstraints}}


\begin{table}[t]
  \begin{center}
  \caption{Bounds on the $\dij{AB}$ as they (a) appear in Ref. \cite{Ciuchini:2007ha} and (b) are modified when the first two generation scalars are $\mathcal{O}(5~{\rm TeV})$. The bounds on LR and RL type insertions are negligible for the models we consider. We have taken the RR/13 bound to be the same as that of the LL/13 bound for simplicity  (see text). \label{ciuchiniconstraints}}
    \subtable[]{\label{ciuchinis}
\begin{tabular}{|| c | c | c ||}
\hline\hline
ij & LL & RR\\
\hline
12 & $6\EE{-4}$ &  $0.09$\\
13 & $0.15$ &  ${\approx 10^{-1}}$\\
23 & $0.12$ &  ${\approx 10^{-1}}$\\
\hline\hline
\end{tabular}}\qquad\qquad
    \subtable[]{\label{ourconstraints}
\begin{tabular}{|| c | c | c ||}
\hline\hline
ij & LL & RR\\
\hline
12 & $0.06$ & $0.09$\\
13 & $0.15$ & $0.15$\\
23 & $-$ & $-$\\
\hline\hline
\end{tabular}}
  \end{center}
\end{table}


Bounds on the $\dij{AB}$ entries for leptonic processes assuming average slepton masses of $500~{\rm GeV}$ may be found in Ref. \cite{Ciuchini:2007ha} and have been reproduced in Table \ref{ciuchinis}. Two of the RR entries lack solid constraints due to cancellations that occur in various regions of the space of MSSM parameters. The orders of magnitude for these bounds are generically expected to be of the same order as the corresponding LL constraint.\footnote{We will be considering $A$ terms that vanish at the messenger scale  ($\mathcal{O}(100~{\rm TeV}$) and become non-zero only through renormalization. The LR entries are therefore small compared to the slepton masses for all $i,j$.}

In models where the first two generation sleptons are of order 5 TeV, the bound on $\dg{LL}{12}$ may be relaxed. To estimate the corresponding LL/12 bound that such a model must satisfy, consider the three relevant processes considered in Ref. \cite{Ciuchini:2007ha}: $\mu\to e\gamma$, $\mu\to eee$, and $\mu\to e$ conversion in $Ti$. The value of $6\EE{-4}$ specifically comes from the decay $\mu\to e \gamma$. A generic Feynman diagram contributing to this decay is illustrated in Figure \ref{feynmangraph}. The branching ratio for this process is approximated by \cite{Hisano:1998fj, Ciuchini:2007ha, Gabbiani:1996hi, Paradisi:2005fk}:
\begin{equation}
\label{branchingfraction}
\frac{BR(l_i\to l_j\gamma)}{BR(l_i\to l_j \nu_i\bar{\nu}_j)} \sim \frac{\alpha^3\delta_{ij}^2}{G_F^2\tilde{m}^4}\tan^2\beta
\end{equation}
where $\tilde{m}$ is the average slepton mass. The constraint is set by requiring that the calculated branching ratio does not exceed the experimental upper bound of $1.2\times 10^{-11}$. Therefore, the bound on $\dg{LL}{12}$ may be safely relaxed by two orders of magnitude to $6\EE{-2}$.\footnote{The other two processes, $\mu\to eee$ and $\mu\to e$ conversion in $Ti$, have branching ratios that are proportional to \eqref{branchingfraction} but smaller by a factor of $\mathcal{O}\(\alpha \)$, whereas the existing experimental constraints are only more stringent by one order of magnitude. (See \cite{Ciuchini:2007ha}). Therefore, the argument is not spoiled by these processes.}

A similar argument would of course apply to the processes $ \tau\to\mu\gamma$ and $\tau\to e\gamma$, indicating that the bounds on the LL/23 and LL/13 entries are weaker as well. Allowing for such large entries, however, pushes us toward the limit of applicability of the mass insertion approximation for which we can trust formulas such as \eqref{branchingfraction} \cite{Paradisi:2005fk}. Furthermore, with all entries of the slepton mass matrices large, we must worry about so-called multimass insertions leading to larger rates for $\mu\to e\gamma$, for example \cite{Hisano:1998fj, Giudice:2008uk}. When models obey the original constraint on the LL/13 entry, $\dg{LL}{13}<0.15$, then an $\ord{1}$ LL/23 entry is not expected to lead to large higher order contributions. We emphasize that a more complete analysis involving numerical calculations and scans over the parameter space is needed to fully examine the model, but this is beyond the scope of this work.\footnote{An analysis such as that performed for the quark sector in Ref. \cite{Giudice:2008uk} could also be done.}


With these considerations in mind, the bounds appropriate when the first two generation sleptons are $\mathcal{O}(5~{\rm TeV})$ are listed in Table \ref{ourconstraints}. Note that we have taken the bound for the RR/13 entry to be identical to that of the LL/13 entry. In this way, we need not worry about any cancellations in the FCNC amplitudes (which tend to relax the bound). For reasons discussed below, this does not provide any significant additional constraint on the model.

\begin{figure}[ht]\label{plot-a}
  \begin{center}
   \includegraphics[scale=0.75]{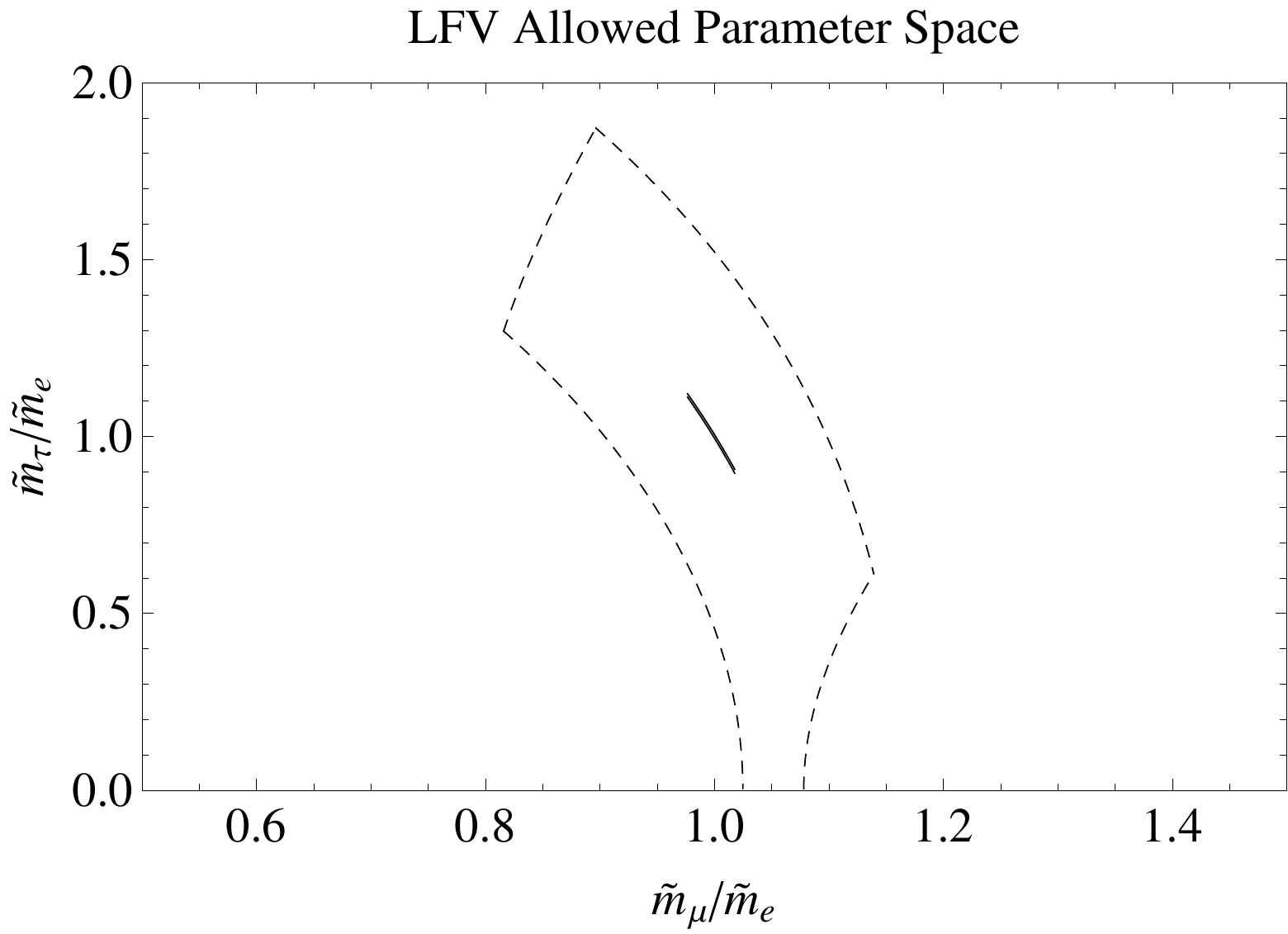}
   \end{center}
   \caption{Mass ratios satisfying LFV constraints. The smaller area (solid outline) corresponds to the constraints of Table \ref{ciuchinis}, while the larger area (dashed outline) corresponding to Table \ref{ourconstraints} is allowed when the first two generation sleptons are $\ord{5~\TeV}$.}
 \end{figure}

In Figure \ref{plot-a} we have provided plots of the allowed parameter space. The smaller area outlined by a solid line in each plot corresponds to the MSSM case where all three generations of sleptons are $\sim 500~\GeV$, whereas the larger area surrounded by a dashed line is allowed when the first two generations are $\mathcal{O}(5~\rm{TeV})$. Interestingly, light third generations sleptons remain viable when the first two generations are nearly equal in mass.



\section{Application: A Single-Sector Dual}\label{dual}

Reference \cite{Gabella:2007cp} considered a gravity dual to single-sector SUSY breaking \cite{ArkaniHamed:1997fq} in five dimensions. The background of the model, which is inspired by the Randall-Sundrum scenario \cite{Randall:1999ee}, is approximately AdS. While AdS is compatible with supersymmetry \cite{Shuster:1999zf}, in this model SUSY is broken by a deformation of the metric in the IR. In particular, the authors of \cite{Gabella:2007cp} consider the following:
\begin{eqnarray}
ds^{2} &=& e^{-2A(z)}\eta_{MN}dx^M dx^N,
\label{metric}\\
e^{-2A(z)} &=& \frac{1}{(kz)^{2}} \[ 1-\epsilon  \(\frac{z}{z_{1}}  \)^{4}  \].
\label{ourmetric}
\end{eqnarray}
The small parameter $\epsilon=0.05$ arises from an underlying 10D supergravity solution and characterizes the size of the AdS deformation in the IR. For our purposes, its value is taken to be freely chosen (but see Appendix D of \cite{Gabella:2007cp}). This deformation lifts the masses of the scalar zero modes, while the fermion masses are protected by chiral symmetry.

The space is compactified on a $\mathbb{Z}_2$ orbifold of radius $R$, with a UV (IR) brane located at $z=z_0$ ($z=z_1$). Choosing the curvature scale $k\sim M_5$, where $M_5$ is the five-dimensional Planck scale then implies $k\simeq 10^{-3/2}m_P=7.7\times 10^{16}$ GeV. The other model parameters are chosen to be:
\begin{eqnarray}
\label{modelparameters}
&&\pi k R = 28.42,\qquad z_1=(k e^{-\pi k R})^{-1}=(35~{\rm TeV})^{-1}.\nonumber
\end{eqnarray}
At this point, we will also specify the ratio of the Higgs VEVs, $\tan\beta=10$.




\subsection{SM Fermions}
Each standard model fermion is identified with the zero mode of a 5D fermion field transforming under the appropriate gauge group.  For each left-handed standard model doublet, we introduce a doublet of 5D fields sharing a common localization parameter, $c_L^i$. For $c_L^i > 1/2$ ($c_L^i < 1/2$), the doublet is UV (IR) localized. Similarly, for each right-handed SM singlet, we introduce a single 5D field with localization parameter $c_R^i$. The familiar localization features for bulk fermions \cite{Gherghetta:2000qt} are retained in this model. Thus, fields are UV (IR) localized for $c_R^i < -1/2$ ($c_R^i > -1/2$).

The zero modes obtain masses through wavefunction overlap with the UV localized Higgs fields, with effective 4D Yukawa couplings, $Y_{\Psi}$, related to 5D bulk Yukawa couplings, $Y_{\Psi}^{5D}$, by expression (2.4) of Ref. \cite{Gabella:2007cp}:
\begin{equation}
\label{yukawafourd}
Y_{\Psi}=Y_{\Psi}^{5D}k\sqrt{\frac{1/2-c_L}{(kz_1)^{1-2c_L}-1}}\sqrt{\frac{1/2+c_R}{(kz_1)^{1+2c_R}-1}}.
\end{equation}

A large range of 4D Yukawa couplings is obtained by varying the parameters $c_L$ and $c_R$ of each field over a range of $\mathcal{O}(1)$. A slight hierarchy in the 5D couplings of $\ord{10^{-3}-10^{-2}}$ is needed to satisfy experimental constraints for FCNCs in quark processes.






\subsection{Scalar Superpartners \& Soft Masses}

The localization parameter, $c_{L/R}$, of each fermion is related to that of the scalar partner, $b$, by supersymmetry as \cite{ Gherghetta:2000qt}:
\begin{equation}
b=\frac{3}{2}\mp c_{L/R}.\nonumber\\
\end{equation}
The localization parameter determines the soft mass of the scalar. In the small $\epsilon$ limit \cite{Gabella:2007cp},
\begin{equation}
\label{scalarmassformula}
\td{m}^2=\epsilon\frac{(1-b)(b+10)}{(kz_1)^4}\frac{(kz_1)^{1+b}-(kz_1)^{1-b}}{(kz_1)^{1-b}-(kz_1)^{b-1}}k^2+\ord{\epsilon^2}.
\end{equation}

For the choices of the parameters \eqref{modelparameters}, scalar masses can be vanishingly small for $b\ll  0$ or $\mathcal{O}(z_1^{-1}=35~{\rm TeV})$ for $b>1$. The size of soft masses is thus tied to the localization of the superfield. Because the Higgs is located on the UV brane, the light (heavy) fermions will be IR (UV) localized. Thus, their superpartners will be heavy (light), because SUSY is broken in the IR.

Note that generically one requires degeneracy in masses between left- and right-handed superpartners to SM fields charged under $U(1)_Y$. Such degeneracy is necessary (in either mixing scenario) in order to avoid a large hypercharge Fayet-Iliopoulos (FI) term after integrating out the heavy scalars \cite{Cohen:1996vb}. Specifically, the requirement is:
\begin{equation}
{\rm Tr}\(Y\td{m}_i^2\)={\rm Tr}\(\td{m}_Q^2+\td{m}_D^2-2\td{m}_U^2-\td{m}_L^2+\td{m}_E^2\)\simeq 0\label{trace}
\end{equation}
where $Y$ is the hypercharge operator.

To determine the physical masses, the soft masses are run down from the messenger scale to the electroweak scale using Softsusy \cite{Allanach:2001kg}. The messenger scale corresponds to the Kaluza-Klein scale, $m_{KK}=110~\TeV$.
\subsection{Tuning the Model}



\begin{table}[htp]
  \begin{center}\caption{Soft and physical masses for sleptons as proposed in Ref. \cite{Gabella:2007cp}.
  \label{tab:masstables}} 
    \subtable[]{
\begin{tabular}{|| c | c ||}
\hline\hline
Sparticles  & Soft Mass (TeV)\\
\hline
$\tilde{e}_{L,R}$, $\tilde{\nu}_{e_L}$ & 10.14\\
\hline
$\tilde{\mu}_{L,R}$, $\tilde{\nu}_{\mu_L}$ & 5.12\\
\hline
$\tilde{\tau}_{L,R}$, $\tilde{\nu}_{\tau_L}$ & 0.468\\
\hline\hline
\end{tabular}\label{masstable-a}}\qquad\qquad
    \subtable[]{
\begin{tabular}{|| c | c ||}
\hline\hline
Sparticles  & Physical Mass (TeV)\\
\hline
$\tilde{e}_{L}$, $\tilde{e}_{R}$, $\tilde{\nu}_{e_L}$ & 10.160, 10.150, 10.160\\
\hline
$\tilde{\mu}_{L}$, $\tilde{\mu}_{R}$, $\tilde{\nu}_{\mu_L}$ & 5.145, 5.130, 5.145\\
\hline
$\tilde{\tau}_{1}$, $\tilde{\tau}_{2}$, $\tilde{\nu}_{\tau_L}$ & 0.511, 0.630, 0.633\\
\hline\hline
\end{tabular}\label{masstable-b}}

  \end{center}
\end{table}

The original model of Ref. \cite{Gabella:2007cp} did not consider LFV and so SM lepton masses were all that were considered in constraining the model parameters, which yielded the spectrum in Table \ref{tab:masstables}. As discussed above, a light third generation slepton requires that the first two generations be nearly degenerate in mass. That is to say, we wish to lie in the modest vertical band of allowed parameter space in the lower part of Figure \ref{plot-a}. Thus, the muon and electron must share similar localizations. Since the effective 4D parameters arise from overlap integrals, we must introduce some hierarchy into the 5D Yukawa couplings. Shifting the left-handed electron multiplet towards the UV so that the selectron obtains a mass of about $5~\TeV$ (instead of $\sim10~{\rm TeV}$) accomplishes the task, requiring only an ($\ord{10^{-2}}$) hierarchy in the 5D lepton Yukawa couplings. The stau and tau sneutrino in this case remain relatively light, allowing for potentially interesting collider phenomenology.

We have modified Softsusy to incorporate the assumption \eqref{assumption}, allowing for a full calculation of the mass spectrum and $\dg{AB}{ij}$'s, including renormalization effects and EWSB but neglecting the (very small) corrections to the RH sneutrino spectrum. In Table \ref{masstableb-a}, we have provided modified localizations and 5D Yukawa couplings for the charged leptons, while Table \ref{physmasstableb-a} contains the corresponding physical masses after RGE. In Table \ref{tab:deltas}, we have listed the resulting mass-insertion parameters, which are easily seen to satisfy the constraints of Section \ref{fcncconstraints}.

\begin{table}[ht]
  \begin{center}
    \caption{Tuned soft and physical masses in the model. The listed value of $c$ determines both localization parameters as $c_L=-c_R=c$.\label{tab:masstablesb}} 
    \subtable[]{\label{masstableb-a}
\begin{tabular}{|| c | c | c | c ||}
\hline\hline
Sparticles  & $Y_{\Psi}^{5D}k$ & c & Soft Mass (TeV)\\
\hline
$\tilde{e}_{L,R}$, $\tilde{\nu}_{e_L}$ & $4.46\EE{-3}$ & 0.470 & 5.00\\
\hline
$\tilde{\mu}_{L,R}$, $\tilde{\nu}_{\mu_L}$ & 1 & 0.467 & 5.12\\
\hline
$\tilde{\tau}_{L,R}$, $\tilde{\nu}_{\tau_L}$ & 1 & 0.601 & 0.468\\
\hline\hline
\end{tabular}}
    \subtable[]{\label{physmasstableb-a}
\begin{tabular}{|| c | c ||}
\hline\hline
Sparticles  & Physical Mass (TeV)\\
\hline
$\tilde{e}_{L}$, $\tilde{e}_{R}$, $\tilde{\nu}_{e_L}$ & 5.030, 5.008, 5.030\\
\hline
$\tilde{\mu}_{L}$, $\tilde{\mu}_{R}$, $\tilde{\nu}_{\mu_L}$ & 5.146, 5.130, 5.146\\
\hline
$\tilde{\tau}_{1}$, $\tilde{\tau}_{2}$, $\tilde{\nu}_{\tau_L}$ & 0.506, 0.635, 0.638\\
\hline\hline
\end{tabular}}
  \end{center}
\end{table}

At this stage, there is no difficulty including right-handed neutrinos and generating tiny Dirac neutrino masses. However, there remains considerable freedom in choosing the various bulk parameters, corresponding to the additional parameters introduced by adding bulk right-handed neutrino mass matrices.

Nevertheless, we can state on general grounds that the bulk right-handed neutrinos must be localized far away from the Higgs into the IR. Therefore, the right-handed sneutrinos are universally heavy, obtaining masses of order $25~{\rm TeV}$ or greater. The RR-type constraints will be automatically satisfied.

\begin{table}[ht]
  \begin{center}
  \caption{$|\dg{AB}{ij}|$'s resulting from the modified spectrum.\label{tab:deltas}}
\begin{tabular}{|| c || c | c | c ||}
\hline\hline
AB/ij & 12 & 23 & 13\\
\hline
\hline
LL & 0.059 & 0.97 & 0.10\\
\hline
RR & 0.059 & 0.97 & 0.10\\
\hline
LR & $3.4\EE{-22}$ & $1.0 \EE{-21}$ & $3.1\EE{-22}$\\
\hline
RL & $1.5\EE{-23}$ & $7.4 \EE{-22}$ & $1.9\EE{-21}$\\
\hline\hline
\end{tabular}

  \end{center}
\end{table}

\section{Conclusion}\label{conclusion}

We discussed leptonic flavor violation in the MSSM when hierarchical soft terms arise from SUSY breaking in an extra dimension. Hierarchical soft terms can arise naturally when both the amount of SUSY breaking and effective 4D Yukawa couplings are tied to field localizations in the extra dimension. We showed in particular that light third generation sfermions are generically allowed if the first two generations remain fairly degenerate in mass and $\mathcal{O}(5~{\rm TeV})$.

We demonstrated the possibility concretely through application to a dual model of single-sector supersymmetry breaking proposed Gabella {\it et al.} \cite{Gabella:2007cp}, showing that moderate adjustments yield a model consistent with flavor physics. The localization mechanism accounts for both the pattern of supersymmetry breaking soft masses as well as most of the hierarchy in the 4D Yukawa couplings.

It is interesting to note that in this class of models, it is generally not possible for $\delta_{12}$ and $\delta_{13}$ to be simultaneously very small. Meanwhile $\delta_{23}$ is generically $\mathcal{O}(1)$. This implies that negative results in future flavor violation searches can be used to set lower bounds on the masses of the first two generation sleptons when the third generation is approximately $500~{\rm GeV}$. Stated differently, it is in principle possible to derive a lower bound for the rate of $\mu\to e\gamma$ in a generic model with first two generation sleptons of approximately $5~{\rm TeV}$ and third generation sleptons of approximately $500~{\rm GeV}$. We leave a detailed analysis of this possibility for future work.

There is ample room to expand our analysis. For example, there are significant uncertainties in the neutrino mixing angles \cite{Maltoni:2004ei}, whereas we have only considered the best-fit values here. Moreover, for simplicity we have used conservative estimates for the bounds on flavor violating amplitudes. A more accurate and in depth analysis similar to that of Ref. \cite{Giudice:2008uk} remains an interesting possibility for future work. Finally, we have considered only one possible realization of neutrino oscillations (i.e., assumption \eqref{assumption}). There is a rich literature discussing how the PMNS matrix can arise from mixing in both the neutrino and charged lepton sectors. It would be interesting to explore these possibilities in greater depth, as they are expected to have non-trivial consequences for models with non-universal soft terms.

\section*{Acknowledgements}
I would like to thank Tony Gherghetta for suggesting this project and Joel Giedt for helpful advice. This work was supported by a grant from the University of Minnesota.

\bibliography{OralPaper}
\bibliographystyle{h-physrev}
\end{document}